\documentclass[preprint]{aastex}
\begin{document}
\newcommand{\be}{\begin{equation}}
\newcommand{\en}{\end{equation}}
\def\ltsima{$\; \buildrel < \over \sim \;$}
\def\lsim{\lower.5ex\hbox{\ltsima}}
\def\loe{\lower.5ex\hbox{\ltsima}}
\def\gtsima{$\; \buildrel > \over \sim \;$}
\def\gsim{\lower.5ex\hbox{\gtsima}}
\def\goe{\lower.5ex\hbox{\gtsima}}
\def\rref{\par\noindent\hangindent=1.5truecm}
\def\aa #1 #2 {A\&A #1 #2}
\def\aass #1 #2 {A\&AS #1 #2}
\def\araa #1 #2 {ARA\&A #1 #2}
\def\mon #1 #2 {MNRAS #1 #2}
\def\apj #1 #2 {ApJ #1 #2}
\def\apjss #1 #2 {ApJS #1 #2}
\def\apjl #1 #2 {ApJ #1 #2}
\def\astrj #1 #2 {AsJ #1 #2}
\def\nat #1 #2 {Nature #1 #2}
\def\pasj #1 #2 {PASJ #1 #2}
\def\pasp #1 #2 {PASP #1 #2}
\def\msai #1 #2 {Mem. Soc. Astron. Ital. #1 #2}
\def\ass #1 #2 {Ap. Sp. Science #1 #2}
\def\sci #1 #2 {Science #1 #2}
\def\phrevl #1 #2 {Phys. Rev. Lett. #1 #2}
\newcommand{\si}{\left(\frac{\sigma}{0.005}\right)}
\newcommand{\psrdot}{\frac{\dot{P}_{-20}}{P_{-3}^3}}
\newcommand{\eee}{$e^{\pm}$}
\newcommand{\ergs}{\rm \ erg \; s^{-1}}
\newcommand{\msol}{\su M_{\odot} }
\newcommand{\etal}{et al.\ }
\newcommand{\Po}{$ P_{orb} \su$}
\newcommand{\pot}{$ \dot{P}_{orb} / P_{orb} \su $}
\newcommand{\myr}{ \su M_{\odot} \su \rm yr^{-1}}
\newcommand{\ppp}{ \dot{P}_{-20} }
\newcommand{\ci}[1]{\cite{#1}}
\newcommand{\bb}[1]{\bibitem{#1}}
\newcommand{\pdot}{ $\dot{P}_{orb}$ \su}
\newcommand{\befl}{ \vspace*{-17pt} \begin{flushright}}
\newcommand{\enfl}{\end{flushright}}
\def\deg {^\circ}
\def\mdot {\dot M}
\def\kms  {\rm \ km \, s^{-1}}
\def\cms  {\rm \ cm \, s^{-1}}
\def\gs   {\rm \ g  \, s^{-1}}
\def\cmtre {\rm \ cm^{-3}}
\def\cmdue {\rm \ cm^{-2}}
\def\gcmdue {\rm \ g \, cm^{-2}}
\def\gcm  {\rm \ g \, cm^{-3}}
\def\rsole {~R_{\odot}}
\def\msole {~M_{\odot}}
\def\fH {{\cal H}}
\def\op {{\cal K}}
\def\nupa{\vfill\eject\noindent}
\def\der#1#2{{d #1 \over d #2}}
\def\inizio{\2acapo\penalty+10000}
\def\fine{\acapo\penalty-10000\blank}
\received{} \accepted{} 
\journalid{}{}
\articleid{}{}

\title{Correlations in the QPO Frequencies of Low Mass X-Ray Binaries 
and the Relativistic Precession Model}  
\author{{\bf Luigi Stella\altaffilmark{1,4}, Mario Vietri\altaffilmark{2}
and Sharon M. Morsink\altaffilmark{3}}} 
\altaffiltext{1}{Osservatorio Astronomico di Roma, Via Frascati 33, 
00040 Monteporzio Catone (Roma), Italy, 
e-mail: stella@coma.mporzio.astro.it} 
\altaffiltext{2}{Universit\`a di Roma 3, Via della Vasca Navale 84, 
00147 Roma, Italy, e-mail: vietri@corelli.fis.uniroma3.it }
\altaffiltext{3}{Department of Physics, University of Wisconsin-Milwaukee, 
P.O. Box 413, Milwaukee, WI 53201
morsink@pauli.phys.uwm.edu }
\altaffiltext{4}{Affiliated to the International Center for Relativistic
Astrophysics}

\begin{abstract}

A remarkable correlation between the centroid frequencies of quasi periodic
oscillations, QPOs, (or peaked noise components) from low mass X-ray binaries,
has been recently discovered 
by Psaltis, Belloni \& van der Klis (1999). This correlation 
extends over nearly 3 decades in frequency and encompasses both 
neutron star and black hole candidate systems. 
We discuss this result in the light of the 
relativistic precession model, which has been proposed to 
interpret the kHz QPOs as well as some of the lower frequency QPOs of 
neutron star low mass X-ray binaries of the Atoll and Z classes. 
Unlike other models the relativistic precession model
does not require the compact object to be a 
neutron star and can be applied to black hole candidates as well. 
We show that the predictions of the relativistic precession model
match both the value and dependence of the correlation 
to a very good accuracy and without resorting to additional assumptions.

\end{abstract}

\keywords{X--ray: stars -- accretion, accretion disks -- relativity -- black 
hole physics -- stars: neutron}
  
\section {Introduction}
 
Old accreting neutron stars, NSs, in low mass X-ray binaries, LMXRBs, display
a complex variety of quasi-periodic oscillation, QPO, modes in their X-ray 
flux. The {\it low frequency} QPOs ($\sim 1-100$~Hz), 
were discovered and studied in the eighties
(for a review see van der Klis 1995). In high luminosity Z-sources  
these QPOs were further classified into horizontal, normal and flaring 
branch oscillations (HBOs, NBOs and FBOs, respectively) depending on the 
instantaneous position occupied by a source in the X-ray colour-colour 
diagram. 
The kHz QPOs (range of $\sim 0.2$ to $\sim 1.25$~kHz) that were
revealed and investigated with RXTE (van der Klis 1998, 1999a and references
therein) in a number of NS LMXRBs 
involve timescales similar to the dynamical timescales close to the NS.
A common phenomenon is the presence of a pair of
kHz QPOs (centroid frequencies of $\nu_1$ and $\nu_2$)
which drift in frequency while mantaining their
difference frequency $\Delta\nu = \nu_2 - \nu_1 \approx 250-350$~Hz
approximately constant. kHz QPOs show remarkably similar properties 
across NS LMXRBs of the Z and Atoll groups, the luminosity 
of which differs by a factor of $\sim 10$ on average.   
During type I bursts from seven Atoll 
sources, a nearly coherent signal at a frequency of 
$\nu_{burst} \sim 350 - 600$~Hz
has also been detected. Four of these sources display also a pair 
of kHz QPOs. While in two of these  $\nu_{burst}$ appears to be 
consistent, to within the errors, with the frequency separation 
of the kHz QPO pair $\Delta\nu$, 
or twice its value $2\Delta\nu$, there are 
currently two sources (4U1636-53, Mendez et al. 1999, and 4U1728-34,  
Mendez \& van der Klis 1999) for which $\nu_{burst}$ is significantly 
different from $\Delta\nu$ and its harmonics. 

The $15-60$~Hz HBO frequency, $\nu_{HBO}$, shows an approximately quadratic
dependence ($\sim \nu_2^2$) on the higher kHz QPO frequency that is observed 
simultaneously in many  Z-sources 
(Stella \& Vietri 1998b; Psaltis et al. 1999). A similar 
$\sim \nu_2^2$ dependence holds also for the centroid frequency of QPOs or 
peaked noise  components of several Atoll sources, suggesting a close
analogy  with HBOs (Stella \& Vietri 1998a; Ford \& van der Klis 1998; Homan et
al. 1998). Evidence for an equivalent to the NBOs/FBOs of Z-sources has
been  found in two Atoll sources (Wijnands et al.
1999; Psaltis, Belloni \& van der Klis 1999, hereafter PBV). 

The origin of these QPOs is still debated. Beat frequency
models, BFMs, require that interactions at two distinct radii take 
place simultaneously. These involve 
disk inhomogeneities at the magnetospheric boundary  
and sonic point radius which are accreted at the beat frequency between the
local Keplerian frequency, $\nu_\phi$, and the NS spin frequency,
$\nu_s$, giving rise to the HBOs (Alpar \& Shaham 1985; Lamb et al. 1985)
and the lower frequency kHz QPOs, at $\nu_1$  (Miller et al. 1998a),
respectively.  The higher frequency kHz QPOs are attributed to the
Keplerian motion  ($\nu_2 = \nu_\phi$) at the sonic point radius. 
Therefore, the frequency separation $\Delta\nu$
yields the NS spin frequency ($\nu_s = \Delta\nu$).
 
Simple BFMs are not applicable to  black hole candidates, BHCs, 
because the {\it no hair theorem} 
excludes the possibility that an offset magnetic field or radiation beam can 
be stably anchored to the BH, as required to produce the beating with the 
disk Keplerian frequency. 

A variety of alternative QPO models has been proposed (for a list 
see PBV). 
In the relativistic precession model, RPM, the QPO signals 
arise from the fundamental frequencies
of the motion of blobs in the vicinity of the NS. 
While the higher frequency kHz QPOs (as in other models) correspond to the 
Keplerian frequency of the innermost disk regions, 
the lower frequency kHz QPOs originate in the 
relativistic periastron precession of (slightly) eccentric orbits and the HBOs 
in the nodal precession of (slightly) tilted orbits in the same regions
(Stella \& Vietri 1998a, 1999).
Within this model both the quadratic dependence 
of the HBO frequency on $\nu_2$ and the decreasing separation 
$\Delta\nu$ for increasing $\nu_2$, observed in several Z and Atoll sources 
are naturally explained, (The near
coincidence of  $\Delta \nu $ and $\nu_{burst}$, or $2\nu_{burst}$,  
in the framework of the RPM will be discussed elsewhere.) We note that the
RPM can be applied to BHCs as well.

In a recent study aimed at classifying the QPOs and peaked noise 
components of NS and BHC LMXRBs 
Psaltis, Belloni \& van der Klis (PBV) identified two components
the frequency of which follows a tight correlation over nearly three decades. 
This correlation (hereafter PBV correlation) involves both NS and BHC LMXRBs 
spanning different classes and a wide range of luminosities 
(see the points in Fig.1). 
In kHz QPO NS systems, these components are the lower frequency 
kHz QPOs, $\nu_1$, and the low frequency, HB or HB-like QPOs, $\nu_{HBO}$.
For BHC systems and lower luminosity NS LMXRBs the correlation 
involves either two QPOs or a QPO and a peaked noise component. In all cases 
the frequency separation is about a decade and an approximate
linear relationship ($\nu_{HBO} \sim \nu_1^{0.95}$) holds. 
Note that the QPO frequencies from the peculiar NS system Cir~X-1 varies
over nearly a  decade while closely following the PBV correlation  
and bridging its low and high frequency ends. 
PBV noted also that $\nu_2$ vs. $\nu_1$ relations of different Atoll
and Z-sources line-up with good accuracy (cf. also Psaltis et al. 1998).
If confirmed, the strong similarity of the QPO (and peaked noise) properties 
across NS and BHC systems that the results of PBV unveiled holds
the potential to tightly constrain theoretical models for the QPO
phenomenon. In this letter we show that the predictions of the RPM accurately 
match the PBV correlation, without resorting to additional assumptions. 

\section {The Relativistic Precession Model}

For the sake of simplicity we consider here blobs moving along 
infinitesimally eccentric and tilted test-particle
geodesics.  In the case of a circular geodesic in the equatorial
plane ($\theta=\pi/2$) of a Kerr black hole
of mass $M$ and specific angular momentum $a$,  
we have for the coordinate frequency measured by a static observer at infinity 
(Bardeen et al. 1972)
\begin{equation}
\nu_\phi=\pm M^{1/2}r^{-3/2} [2\pi (1\pm a M^{1/2}r^{-3/2})]^{-1} \ \  
\end{equation}
(we use units such that $G=c=1$). 
$\nu_\phi$ deviates from its Keplerian value at small 
radii. The upper sign refers to prograde orbits.  
If we slightly perturb a circular orbit introducing velocity components in
the $r$ and $\theta$ directions, the coordinate frequencies of 
the small amplitude oscillations within the plane (the epicyclic
frequency $\nu_r$) and in the perpendicular direction (the vertical
frequency $\nu_\theta$) are given by (Okazaki, Kato \& Fukue 1987;
Kato 1990)
\begin{equation}
\nu_r^2=\nu_\phi^2 (1-6M r^{-1} \pm 8aM^{1/2} r^{-3/2}- 3a^2 r^{-2}) \ \ ,
\end{equation}
\begin{equation}
\nu_\theta^2=\nu_\phi^2 (1\mp 4 aM^{1/2}r^{-3/2}+3a^2r^{-2}) \ \ .
\end{equation}
In the Schwarzschild limit ($a=0$) $\nu_\theta$ coincides with 
$\nu_\phi$, so that the nodal precession frequency 
$\nu_{nod} \equiv \nu_\phi - \nu_\theta$ is identically  zero. 
$\nu_r$, on the other hand, is always lower than the other two 
frequencies, reaching
a maximum for $r=8M$ and going to zero at $r_{ms}=6M$, the radius 
of the marginally stable orbit.
This qualitative behaviour of the epyciclic frequency is preserved in the 
Kerr field ($a \neq 0$). Therefore the periastron precession frequency
$\nu_{per}\equiv \nu_\phi-\nu_r$ is dominated by a ``Schwarzschild" term 
over a wide range of parameters (cf. Stella \& Vietri 1999). 

In the RPM the higher and lower frequency kHz QPOs are identified
with $\nu_2=\nu_\phi$ and $\nu_1=\nu_{per}$, respectively, whereas the 
HBOs at $\nu_{HBO}$ are identified with the 2nd harmonics of $\nu_{nod}$
(cf. Stella \& Vietri 1998a, 1999). In fact in 
those few Atoll LMXRBs in which $\nu_{s}$ can be inferred from 
$\nu_{burst}$,  using $\nu_{HBO} = 2\nu_{nod}$ (instead of 
$\nu_{nod}$) provides a fairly good match (Morsink \&
Stella 1999). 
The geometry of a tilted inner disk 
might be such that a stronger signal is produced at 
the even harmonics of the nodal precession 
frequency (e.g. Psaltis et al. 1999). 
By analogy, we assume that also in BHCs $\nu_{HBO}\simeq 2\nu_{nod}$. 
It should be noticed that, within the RPM, $\nu_{s}$ 
can be inferred only indirectly (i.e., by fitting $\nu_{HBO}$) for those NS 
systems that do not display 
coherent pulsations or burst oscillations. Overall the RPM yields a wider 
distribution of spin frequencies than inferred in BFMs based on $\Delta\nu$
(cf. Stella et al. 1999). A magnetosphere is not 
required and, if present, must be such that the motion of the blobs is 
perturbed only marginally,  
implying magnetic fields $\leq 10^8-10^9$~G. 

Fig.1A shows $2\nu_{nod}$ and $\nu_\phi$ as a function of $\nu_{per}$ for 
corotating orbits and selected values of $M$ and $a/M$. 
The high frequency end of each line is dictated by 
the orbital radius reaching $r_{ms}$ (where $\nu_r=0$,  
{\it i.e.} $\nu_\phi = \nu_{per}$). 
It is apparent that $r_{ms}$ depends mainly on $M$ and to a lesser 
extent on $a/M$. 

The separation of the lines in Fig.~1A testifies that while $\nu_{nod}$
depends weakly on the  mass and more strongly on $a/M$, the opposite is 
true for $\nu_\phi$. By taking the
weak field ($M/r \ll 1$) and  slow rotation ($a/M \ll 1$) limit of Eqs. (1)-(3)
the relevant first order dependence is made explicit 
(we use here $m=M/M_{\odot}$) 
\begin{equation}
\nu_\phi \simeq (2\pi)^{-2/5}3^{-3/5} M^{-2/5} \nu_{per}^{3/5} 
\simeq 33\ m^{-2/5}\nu_{per}^{3/5} \ \ {\rm Hz} \ \ , 
\end{equation}
\begin{equation}
\nu_{nod}\simeq(2/3)^{6/5} \pi^{1/5} (a/M) M^{1/5} \nu_{per}^{6/5}
\simeq 6.7 \times 10^{-2}\ (a/M) m^{1/5} \nu_{per}^{6/5} \ \ {\rm Hz} \ \ .
\end{equation}

In the case of rotating NSs the stellar oblateness introduces 
substantial modifications relative to a Kerr spacetime. Approximate 
expressions have been worked out to calculate the precession frequencies arising
from the quadrupole component of the NS field (cf. Morsink \&
Stella 1999; Stella \& Vietri 1999). 
These approximations, however, break down for NSs
spinning within a factor of $\sim 2$ from breakup. 
In order to investigate a wide range of NS spin frequencies, we 
adopted a numerical approach and computed the spacetime metric of the NS 
using Stergioulas' (1995) code, an equivalent of that of Cook et al.
(1992) (cf. also Stergioulas and Friedman 1995).
$\nu_{nod}$ was calculated according to the prescription 
of Morsink \& Stella (1999), whereas $\nu_{per}$ 
was derived from the epicyclic frequency $4\pi^2\nu_r^2 \equiv
d^2V_{eff}/dr^2$, with $V_{eff}$ the effective potential.

Numerical results are shown in Fig.1B for a NS  mass of 
1.95~M$_{\odot}$, a relatively stiff equation of state (EOS AU cf. 
Wiringa et al. 1988) 
and $\nu_{s} = $~300, 600, 900 and 1200~Hz (corresponding to 
$a/M =$~0.11, 0.22, 0.34 and 0.47, respectively).  
While also in this case  $\nu_{\phi}$
depends only very weakly on $\nu_{s}$, the approximately linear dependence of
$\nu_{nod}$ on $\nu_{s}$ is apparent. 
Note that the approximate scalings in Eqs. (4)-(5) 
remain valid over a wide range of parameters. Only for 
the largest values of $\nu_{per}$ and $\nu_{s}$,   
$\nu_{nod}$ departs substantially from the $\sim \nu_{per}^{6/5}$ dependence. 

The measured QPO and peaked noise frequencies giving rise to the PBV 
correlation, together with the higher kHz QPO frequencies from NS systems
($\nu_2$), are also plotted in Fig.1B, to allow for a comparison 
with the model predictions.  
The agreement over the range of frequencies spanned by each kHz QPO NS
system should not be surprising: together with the accurate matching of
the  corresponding $\nu_1-\nu_2$ 
relationship in Z-sources, this is indeed part of the evidence 
on which the RPM model was proposed (Stella \& Vietri 1998a, 1999). 
However the fact that the dependence of $\nu_{nod}$  on $\nu_{per}$ matches 
the observed $\nu_{HBO} - \nu_{1}$ correlation to a good accuracy 
over $\sim 3$ decades in frequency (down to 
$\nu_1$ of a few Hz), encompassing both NS and BHC systems, is new and 
provides additional support in favor of the RPM. 
The observed variation of $\nu_{HBO}$ and $\nu_1$ in individual sources 
(Cir~X-1 is the most striking example) 
provide further evidence in favor of the scaling predicted by the RPM. 
The agreement of $\nu_{\phi}$ and $\nu_{per}$
with the observed $\nu_{2} - \nu_{1}$ relation 
is also accurate. 

In the case of NS LMXRBs, a good matching is obtained 
for relatively stiff EsOS ({\it e.g.} AU as in Fig.1 and UU), which can
also achieve the relatively high masses ($m\sim 1.8-2.0$) that are
required to match the observed $\nu_1$ and $\nu_2$ values
(on the contrary soft or very stiff EsOS do not produce acceptable results). 
For EOS AU and $m=1.95$, the $\nu_{HBO} - \nu_{1}$ values of most NS LMXRBs 
are best matched for $\nu_{s}$ in the $\sim 600$ to 900~Hz range. 
It is apparent from Fig.1b that $\nu_{s}$ as low as $\sim 300$~Hz are 
required for a few Atoll sources (notably 4U1608-52 and 4U1728-34). 
The values above are close to the range of $\nu_s \sim 350-600$~Hz  
inferred from $\nu_{burst}$ in a number of Atoll sources (van der Klis 1999). 
Z-type LMXRBs, appear to require $\nu_{s}$ in the $\sim 600$ to 900~Hz 
range, a possibility that is still open since 
for none of these sources there exists yet a measurement of 
$\nu_{s}$ based on burst oscillations or coherent pulsations
(cf. also Stella, Vietri \& Morsink 1999). 

A word of caution is in order concerning the Atoll source 4U1728-34, which 
displays two distinct branches, one similar to that 
of Z-sources, the other characterised by $\nu_{HBO}$ a factor of $\sim 2$ lower, 
and flips from one branch to the other (Ford \& van der Klis 1998). 
(Only the lower branch is plotted in Fig.1, as in the 
upper branch only the QPOs at $\nu_{HBO}$ and $\nu_2$ have been 
observed so far). While this behaviour may be peculiar to  
this source, still it suggests that, if the lower branch corresponds to 
$\nu_{HBO}\simeq 2\nu_{nod}$ (in agreement with the $\nu_s \sim 360$~Hz
inferred from $\nu_{burst}$ and the idea that only even
harmonics of the  nodal precession signal are generated), 
then for the upper branch, $\nu_{HBO}\simeq 4 \nu_{nod}$. 
One could further speculate that
the same holds for sources closely following the 
PBV correlation, Z-sources in particular, such that their $\nu_s$ might 
be expected in the $\sim 300-600$~Hz range.

For BHC LMXRBs the scatter around the PBV correlation
implies values of $a/M$ in the $\sim 0.1-0.4$ range (see Fig.1A). Note that 
so far no evidence has been found for a BHC QPO (or peaked noise) signal 
that is the equivalent to the higher kHz QPOs (at $\nu_2$) of NS
systems. The points from XTE~J1550-564 while inconsistent with any single
value of $a/M$, might lie along two distinct branches, separated by a factor
of $\sim 2$ in $\nu_{HBO}$, similar to the case of 4U1728-34. 

\section{Discussion} 

Interpreting the PBV correlation within the RPM implies that both NS and BHC
systems share relatively high values of their specific angular momentum 
($a/M\sim 0.1-0.3$ for BHCs or $\nu_s = 300-900$~Hz for relatively stiff EOS
NSs).  
An interesting feature of the $2\nu_{nod}$ vs. $\nu_{per}$ curves 
obtained from relativistic rotating NS models (see Fig.~1b), is that 
for high values of $\nu_{per}$ the 
increase of $\nu_{nod}$ for increasing $\nu_{s}$ tends to saturate,  
such that the curves for relatively high values of $\nu_{s}$ are 
closely spaced. This results from the increasingly important role of 
the (retrograde) quadrupole nodal precession relative to the 
(prograde) frame dragging nodal precession. 
This might help explain the relatively narrow range of 
$\nu_{HBO}$ in Z-sources, if these possess    
$\nu_{s}$ in excess of $\sim 500-600$~Hz. Population synthesis calculations 
show that a significant fraction of low magnetic field, relatively stiff EOS
NSs in LMXRBs can be spunup by accretion torques to $\nu_{s} \sim 1$~kHz, 
provided that $> 0.3-0.4$~M$_{\odot}$ are accreted (Burderi et al. 1999).
The combination of the relatively high NS mass and spin frequency required
by  the RPM is at least qualitatively in agreement with evolutionary
scenarios for LMXRBs. Detecting the lower frequency end of the
$\nu_2-\nu_1$  correlation in NS systems would provide an important test of
the  RPM interpretation. 

In the Kerr metric, the increase in $\nu_{nod}$ for increasing 
values of $a/M$ shows no signs of saturation (Fig.1A). 
BHCs can in principle achieve extreme values 
of $a/M \simeq 1$ (NS are instead limited to values of 
$< 0.6-0.7$, cf. Salgado et al. 1994). 
In practice spinup to high values of $a/M$ is unlikely to occur in BHC 
LMXRBs, because of the limited mass 
that can be accreted from the donor star (King \& Kolb 1999). 
In the RPM the higher frequency BHC QPOs 
({\it e.g.} the $\sim 300$~Hz QPOs of GRO1655-40) 
are interpreted in terms of $\nu_1 = \nu_{per}$ 
(while the lower frequency QPOs $\nu_{HBO}=\nu_{nod}$), implying   
$a/M \sim 0.1-0.3$, in agreement with 
accretion-driven spinup scenarios. This is at variance with the high values of 
$a/M$ ($\sim 0.95$ for GRO 1655-40) derived from the $\nu_1 = \nu_{nod}$ 
interpretation of Cui et al. (1998). 
The detection of a QPO signal at $\nu_\phi$ in BHC systems 
({\it i.e.} the equivalent to the higher kHz QPO at $\nu_2$ in NS systems)
could provide an additional test of the RPM. 
Due to the $\nu_\phi \propto M^{-2/5}$ scaling (cf. Eq.4 and Fig.1A) 
the frequency of these BHC QPOs should be somewhat lower than the $\nu_2$ 
of NS systems. The case of the $\sim 300$~Hz QPOs from GRO~1655-40
is especially interesting in this respect: from Fig.1A it is apparent that 
the relevant points lie close to the high frequency end of the 
$a/M=0.1$, $m=7$ line. Since the mass of the BHC in GRO~1655-40 determined 
through optical observations is $\sim 7$~M$_{\odot}$ 
(Orosz \& Bailyn 1997; Shahbaz et al. 1999), 
we conclude that, according to the RPM, $\nu_{per}\sim 300$~Hz 
close to the marginally stable orbit, where
$\nu_{\phi} \sim \nu_{per}$. This suggests that an additional QPO 
signal at $\nu_2 = \nu_{\phi}$ might well be close to  
or even blended with the QPO peak at $\nu_1$. 
The detection of two closeby or even partially overlapping QPO peaks 
close to $\sim 300$~Hz in GRO~1655-40 would therefore provide 
further evidence in favor of the RPM interpretation.
Similar considerations might apply to the $\sim 284$~Hz QPOs 
of XTE~J1550-564 (Homan et al. 1999), 
provided its BHC mass is in the $\sim 7$~M$_{\odot}$
range. 

In the RPM the QPO signals at $\nu_{nod}$ and $\nu_{per}$ 
are produced over a limited range of radii. For kHz QPO NS systems these 
radii are remarkably close to the marginally stable orbit. 
In the model discussed here the eccentricity and tilt angle 
are allowed only very small values,
such that the QPO frequency variations of a given source 
are ascribed entirely to variations of the radius at which the signals 
are produced. 
As shown in section 2, this is sufficient to
interpret the salient features of the $\nu_{HBO}$ and  $\nu_2$ vs. $\nu_1$
correlations.  
Corrections for non-negligible eccentricities or tilt angles are 
small. In the weak field and slow rotation limit, $\nu_{\phi}$, 
$\nu_{per}$ and $\nu_{nod}$ are 
independent of the tilt angle, while a finite eccentricity $e$ would 
give rise to a factor of $(1-e^2)^{3/5}$ and  $(1-e^2)^{-3/10}$ 
on the right hand side of  Eqs. 4 and  5 , respectively. The corresponding
corrections amount to $<16$\% for $e<0.5$.

The orbital radius is 
approximately given by $ r/M \simeq 100\ m^{-2/5}\nu_{per}^{-2/5}$. 
To interpret in this context the lowest observed values of 
$\nu_1$, {\it i.e.} $\sim 2$ and $10$~Hz in NS and BHC systems respectively, 
radii as large as $r/M \sim 30 $ are required.  
While $\sim 4-5$ times larger than the marginally stable orbit and/or 
the NS, these radii are still small enough that the 
gravitational energy released locally can account for the 
observed QPO amplitudes. 
At present we can only speculate on the physical mechanism determining
the radius at which the QPOs are produced and its variation in each  source 
and across different sources. This radius  
must decrease for increasing mass accretion rates, 
as there is much evidence that in Atoll and Z sources the 
frequency of the kHz QPOs is a reliable indicator of the mass accretion 
rate.
Many NS and BHC LMXRBs display two
component  X-ray spectra consisting of a soft 
thermal component, usually interpreted in terms of emission from an optically
thick accretion disk, and a harder often power-law like component, that 
might originate in a hot innner disk region. As the {\it rms} amplitude of 
kHz QPOs and HBOs increases steeply with energy, QPOs are likely associated to 
the hard spectral component.
One possibility is that the QPOs originate at the transition radius 
between the optically thick disk to the hot inner region, perhaps 
as a result of occultation by orbiting blobs. 
A further exploration of this idea is beyond the purpose of this {\it Letter}. 
We note, however, that 
variations of the inner radius of the optically
thick disk have been inferred in a few NS 
and BHC systems (notably  GRS~1915+105, Belloni et al. 1997; see also 
the case of 4U~0614+091, Ford et al. 1997) 
through spectral variability analysis. 
The resonant blob z-oscillations discussed by Vietri \& 
Stella (1998) do occur at well defined radii, but are not relevant to BHCs 
as they require an offset magnetic field anchored to the accreting star. 

In summary the RPM, unlike other models ({\it e.g.} BFMs), 
holds for BHCs as well as NSs; its applications to the striking correlation in
the QPO (and/or peaked noise) frequencies of NS and BHC LMXRBs gives remarkably
good  results. 
 
LS acknowledges useful discussions with T. Belloni, D. Psaltis and 
M. van der Klis.

\newpage

\section*{Figure Caption} 

\figcaption[]{Twice the nodal precession frequency, $2\nu_{nod}$, 
and $\phi$-frequency, $\nu_{\phi}$, 
vs. periastron precession frequency, $\nu_{per}$, for 
black hole candidates of various masses and angular momenta (panel A) 
and rotating neutron star models (EOS AU, $m=1.95$) with selected 
spin frequencies (panel B). 
The measured QPO (or peaked noise) frequencies $\nu_1$, $\nu_2$ and 
$\nu_{HBO}$ 
giving rise to the PBV correlation are also shown in panel B for both 
BHC and NS LMXRBs and in panel A for BHC LMXRBs only; errors bars are 
not plotted (see PBV for a complete list of references). 
Only those cases in which a signal at $\nu_1$ was unambiguously
detected are included. For the sake of clarity NBOs and FBOs are not plotted. 
Filled squares in panel B correspond to the Atoll-source 4U1728-34. 
\label{Figure 1}}

\end{document}